# Incentive Schemes for Mobile Peer–to-Peer Systems and Free Riding Problem: A Survey

Rahul Mishra
M.Tech (Data Engineering) Student
Indraprastha Institute of Information Technology, Delhi (IIIT-D), INDIA.
Phone: +91-9654543666, 9839751235
Email: rahul1128@iiitd.ac.in

Under Supervision of

Dr. Anirban Mondal

## *Abstract*

Mobile peer-to-peer networks are quite prevalent and popular now days due to advent of business scenarios where all the services are going mobile like whether it's to find good restaurants, healthy diet books making friends, job-hunting, real state info or cab-sharing etc. As the mobile users are increasing day by day, peer-to-peer networks getting bigger and complex. In contrast to client server system in peer-to-peer network resource sharing is done on the basis of mutual consent and agreed policies with no central authority and controlling entity. Incentive schemes for P2P networks are devised to encourage the participation and to adhere the policies agreed. P2P services based only on altruistic behaviour of users are facing serious challenges like "**Free riding**" or "**The tragedy of commons**". Free riders are the users who consume the bandwidth of the system (perform downloading) but don't show altruistic behaviour (deny uploading) and act as a parasite for the P2P network. To counter the free riding issue many Incentive schemes are suggested by the researchers. In this paper we will survey the different incentive schemes, their architectures keeping eye on how they handle the challenges of modern P2P network.

# Introduction

Since the advent of first p2p network "Napster" many such networks have come up with improved performance and services such as "Gnutella" and "FreeNet" but in the end they all are facing challenges like free riders and violation of copyright laws. Ensuring and stimulating cooperation among users is prime concern if we want to run any p2p network smoothly but doing so is quite difficult stuff since in p2p sharing networks, we don't have central monitory system or statistics analysis repository.

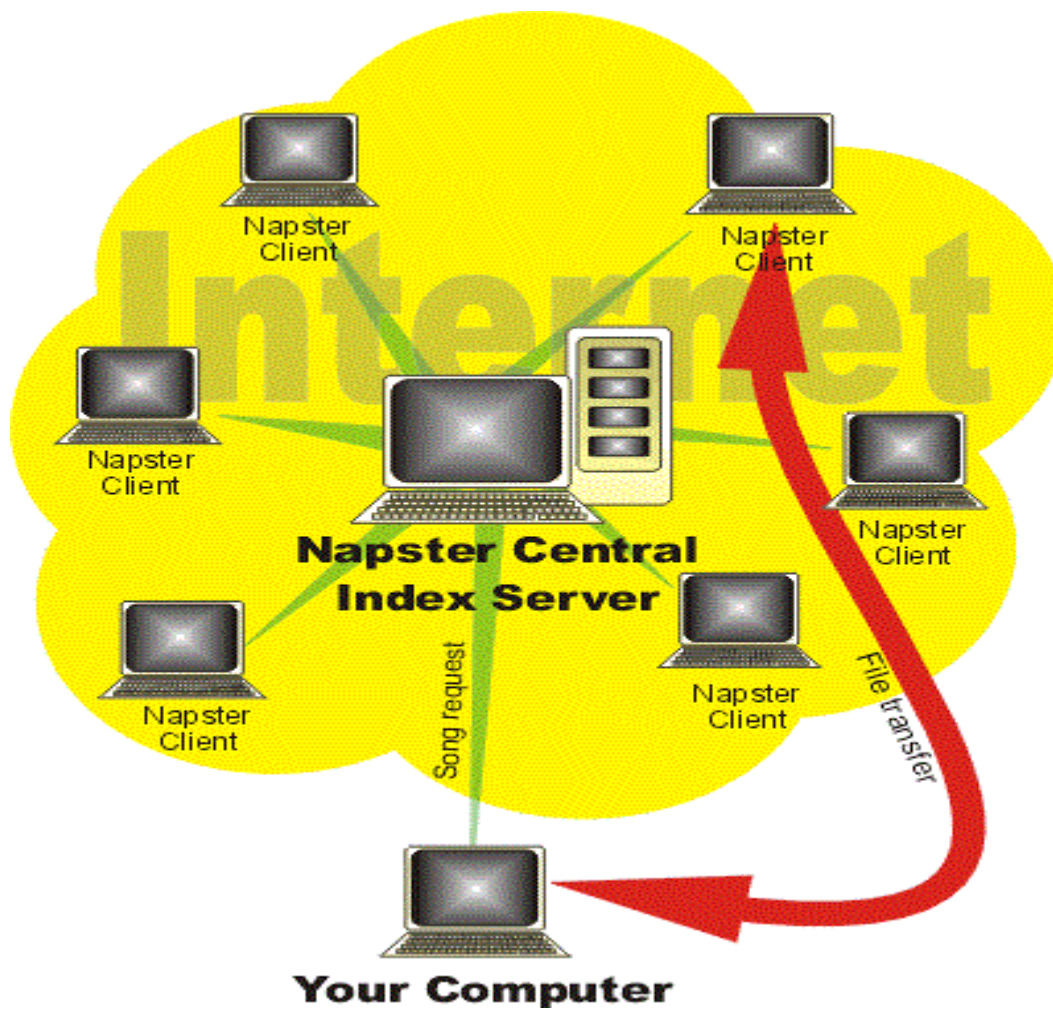

**Figure-1 Basic Napster Architecture**

Maintaining such central authority is in contrast to the spirit of P2P paradigm where users enjoy the unmatchable level of openness, freedom and anonymity. In such environment where no one is keeping track of your activities, its obvious tendency of users to deny cooperation by not sharing their resources while consuming network's scarce bandwidth at the same time. It has been observed that in case of popular p2p network "Gnutella" almost 70% of users don't share anything while more than 50% of the sharing is done by top 1% of users [Eytan Adar and Bernardo A. Huberman]. This means 70% of the user are doing no good to the system and still they getting benefitted through it. System is totally relying on the altruistic behaviour of the top 1-2% of the hosts and as the free riders increase it degrades the quality of services since the search horizon becomes over crowed and many requests are failed to be responded by the system as request message crosses it's time to live before reaching appropriate hosts. Like in Gnutella as we have seen most of the responses are given by few top hosts, due to increase in free riders at some point of time these top hosts will get overwhelmed by requests and saturated to service further.

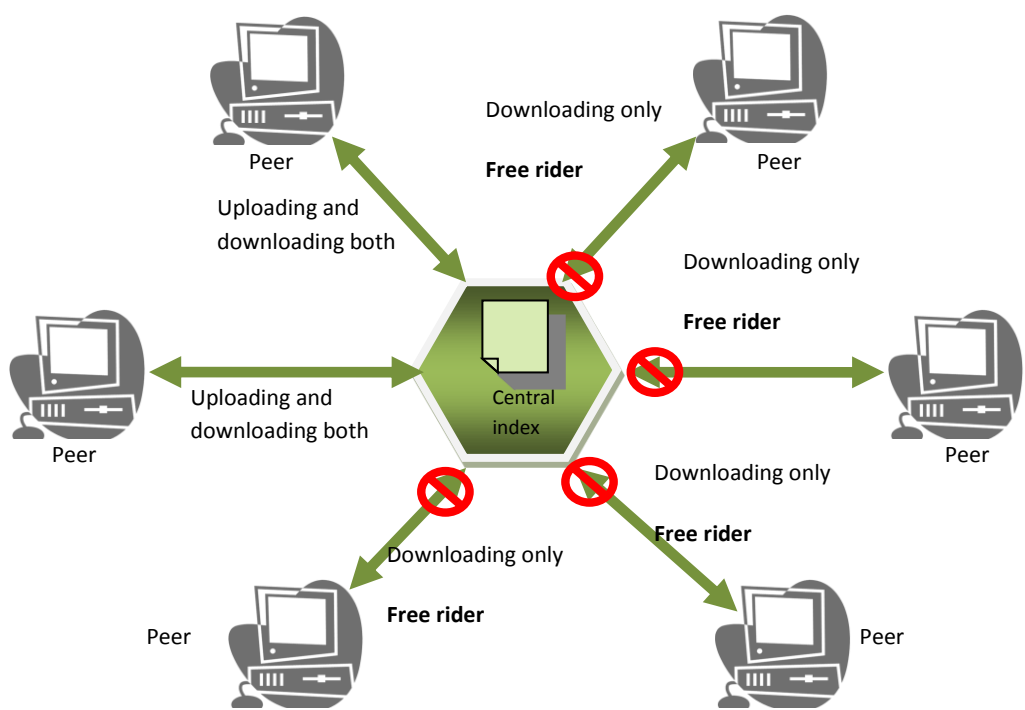

**Figure-2 Free Riding In Currently Prevalent P2P Networks**

As depicted in **Figure-2** most of the peers don't share but download only that's why performance and content availability decreases. To get rid of free riding in Napster, by default, downloads all files into a shared upload directory. In this way when a user downloads a file it is automatically shared.

To encourage the cooperation and reduce the impact of free riders in p2p environment incentive schemes are devised for p2p sharing networks. These schemes distribute rewards in terms of money (micro payments) or usage coupons etc. to the users who cooperate by sharing their resources positively. Alternatively rather than providing incentives to users who are not free riders, we can provide them better *quality of service* (*QoS*). There are so many such incentive schemes are suggested for p2p systems such as game theoretic incentive distribution, reputation based incentive schemes (Eigen trust), Asymmetric incentive schemes and auction based approaches etc.

## *Game-theoretic P2P file-sharing incentive model*

In this approach rather than relying on altruistic behaviour of hosts system provides incentives to the users based on their sharing and downloads, users are allowed to choose their level of sharing and downloading. Users try to maximize their rewards by constructing a game theoretic approach and by managing to get incentives. In this technique resource bidding is performed, peer nodes compete for resource allocation as game theory architecture.

In this scheme it's quite important to identify the files shared by the users to pay appropriate royalties to them and to check the false claims of sharing if any.

**Sharing:** It's totally up to user to select the level of sharing, he is convenient with. There are different levels of sharing mode like No sharing, moderate and all shared.

**Downloading:** Similar to sharing levels different downloading modes are there like no downloads, moderate, heavy downloading.

**Agent Utility Functions:** Agent utility functions govern the choices and preferences of users for different scenarios. Agent utility varies based on some factors like.

- **Amount to Download (AD)**   (Desirable for agents)
- **Network Variety (NV)**   (More network choices are desirable)
- **Disk Space Used (DS)**   (Cost imposed should be less)
- **Bandwidth Used (BW)**   (High upload bandwidth is not desirable)
- **Altruism (AL)**   (Adds utility through satisfaction caused by altruism )
- **Financial Transfer (FT)**   (Agents can spend or earn money by downloading or sharing )

Now we can state the utility function as linear combination of these factors.

$$U_i = f_i^{AD}(AD) + f_i^{NV}(NV) + f_i^{DS}(DS) + f_i^{BW}(BW) + f_i^{AL}(AL) + FT$$

Now each user can draw his strategy of usage based his need and resources available. It can be the any combination of these sharing and downloading levels such as no sharing-heavy downloads , moderate sharing-no downloads etc. Each such strategy leads to appropriate agent utility positive or negative. Agent utility is the probabilistic consequences of their choice of strategy and it depends on various factors such as download size, bandwidth used, incentives, network options, disk space etc. each user tries to maximize his utility positively and the system enters in an equilibrium.

**Equilibria:** Agents behave rationally during participation as they will try to maximize their utility. A weak Nash equilibrium is achieved when no agent can gain utility by changing his strategy, provided that all other agents' strategies are fixed. A strong Nash equilibrium is achieved when every agent would be strictly worse off if he were to change his strategy, given that all other agents' strategies are fixed.

**Micro payments:** Server keep tracks the downloading count and sharing count of each user, as any file transfer comes to an end counters for related users are modified accordingly. After end of each time period users are charged with amount C =f(download - shared). Since a user is charged by same amount for downloading a file as the host given incentive for sharing therefore overall sum of all micro payments remains zero.

## *Reputation based incentive model: Eigen Trust*

In peer to peer scenario incentive schemes are incorporated to increase cooperation and sharing but these mechanisms don't ensure to avoid false claims and inauthentic files uploaded just to get the benefit of uploading provided by the system. These are crap files not having any relevant materials as it claims to have. To check this kind of issue we have to have some heuristics of user's historical uploads and whether the downloader was satisfied or not. To do so we have to maintain reputation of each user to ensure his trustworthiness. Having such global trust value malicious users can be identified and blocked.

In this mechanism each host is assigned with a global trust value which is calculated on the basis of his past sharing and all the users participate in this calculation process.

***Reputation System:*** Each peer records counts of the satisfactory or unsatisfactory experience with each peer it interacts. Global trust value of a peer is calculated based on all records provided by each peer he interacted till now.

*Trust value = sum of all satisfactory – sum of all unsatisfactory*

The real challenge here is not to calculate local trust values but to aggregate them in distributed environment not having any central control to get global trust view, problems that can arise due to this are either the whole local trust values are not considered or network is flooded with trust request messages.

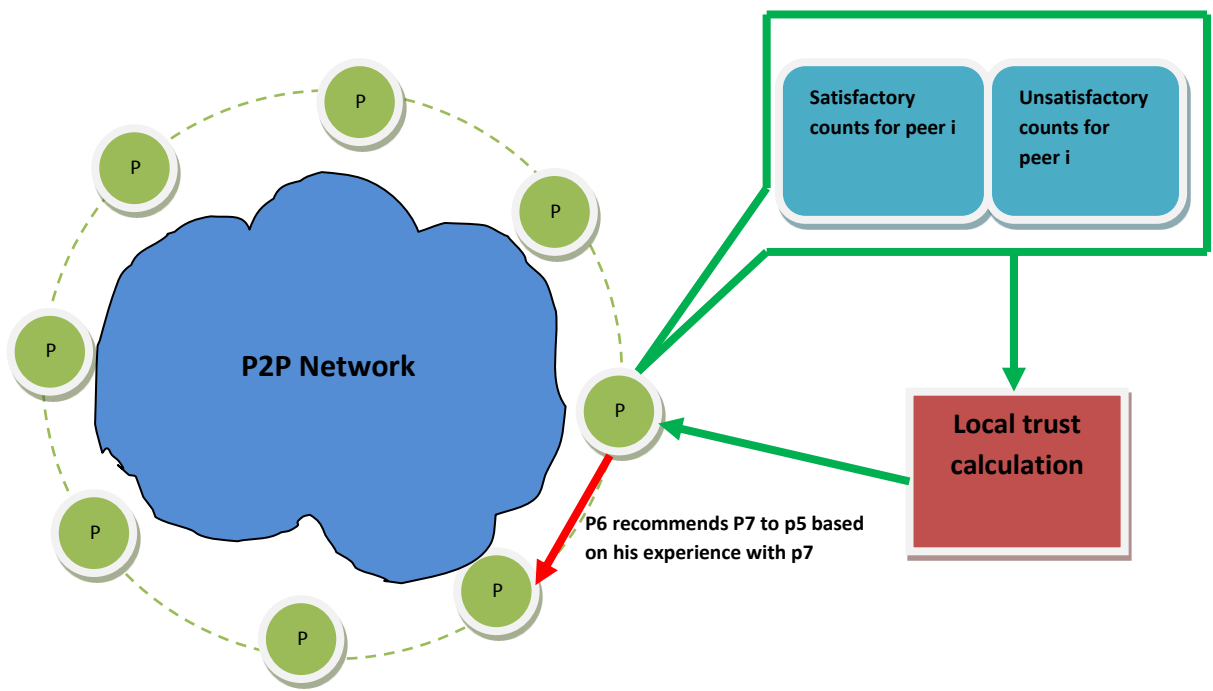

**Figure-3 Eigen Trust Architecture**

***Basic Eigen Trust:*** Solution for issues stated above is Eigen Trust, in this scheme each peer 'i' asks his friends or trusted neighbours ( based on local trust heuristics) to suggest the appropriate peer to whom they themselves trust therefore peer "i" got trust values of his neighbours of neighbours, similarly "i" doing the same steps "i" gets the values of neighbours of neighbours of neighbours and so on.

Like in **Figure-3** peer p6 trusts peer p7 based on his previous experience with it (based on satisfactory and unsatisfactory counts) likewise peer p5 trusts on peer p6. Now peer p6 recommends peer p7 to peer p5 for interaction as it trust p7 and in this way eventually peer p5 starts trusting peer p7. In this way each peer can calculate the global trust values for each other peer by doing so and without flooding the network with trust value messages.

***Issues:*** Issues that can arise in this approach is there is fair chances of malicious users assign high trust value to each other forming an alliance but we can avoid such scenarios if we have pre-trusted peers already in the system distributed in whole network.

***Isolating Malicious Peers:*** Simple and intuitive strategy to avoid malicious peers is to download from most trusted peers only but this strategy is not that promising as it looks because due to this there will be huge overload on most trusted peers to serve every one and system will collapse at some point of time and this scheme will also refrain new peers to build reputation. The solution for such scenario could be making download decision probabilistically on trust values of peers. This scheme will not only balance the load but also will enable the newcomers to attain reputation.

***Incenting Free riders to Share:*** The peers who are most trusted are rewarded by incentives. This will not only motivate free riders to start sharing but also it will boost peers to share only authentic files and to delete inauthentic ones so that high global trust value can be achieved.

## *Auction based incentive scheme in packet forwarding service*

In wireless network environment like mobile ad-hoc networks users are used to have limited resources such as CPU power, memory space and battery etc, therefore they don't want to share these resources altruistically but they rational enough to seek their benefits, Therefore to increase the cooperation quality of service incentives should be given in exchange of services. In this scheme an auction based incentive model is proposed where each router gets paid for packet forwarding.

***Basic algorithm:*** We can define auction based incentive scheme for packet forwarding as follows.

***Flows and Bids:*** At each router an auction takes place (Auction market) and all the packet flows passing through it work as bidders for this router. Bids for each flow are set depending on utility of units of bandwidth achieved against it. Utility function for each user is linear curve and slope is equal to respective bid as shown in **Figure-4**.

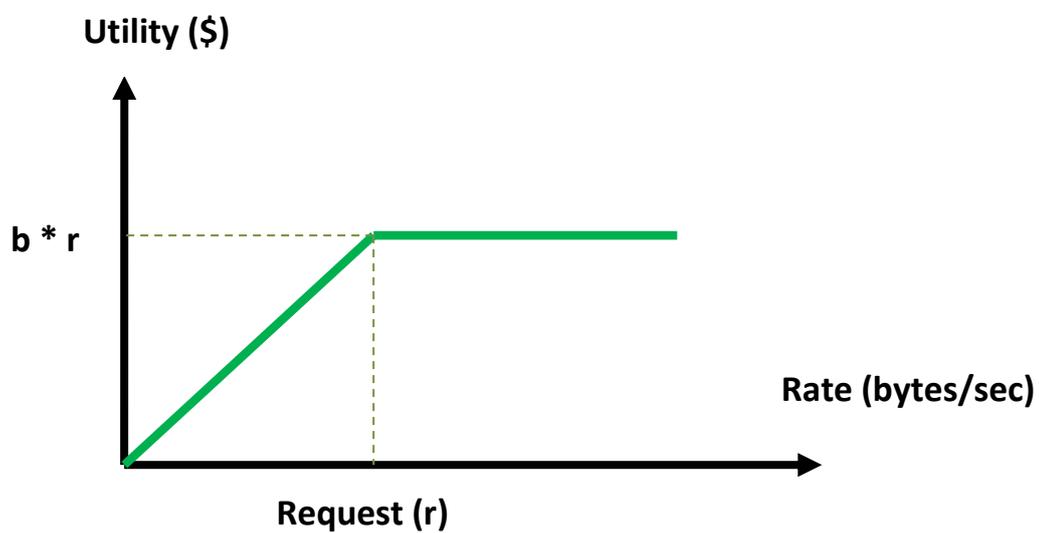

Figure-4

***Router's auction rule:*** Although this scheme uses generalized vickrey auction but with some modification, which is required due to mismatch between user's bid for bandwidth and effort by router to provide that bandwidth. We have to consider channel time also for calculating bids for flow. Bid modified as

$$b' = b \times BW$$

Where b' is modified bid for particular channel time and BW is effective link bandwidth.

***Budget Control and Currency Exchange:*** System assumes that each user is having some amount of initial money before participation in system and he can earn more money by packet forwarding or can spend money using other user's services (forwarding his own packets through others). Therefore each user can assess his wealth status during his participation and use some budget control policies such as requesting lesser bandwidth or limiting the session duration.

***Router's Reserve Price:*** Reserve price for a router constitutes its operating cost, cost of facility, equipment and staffing etc. Therefore based on market research of utilities of different users routers can set their reservation cost to compensate for providing packet forwarding service.

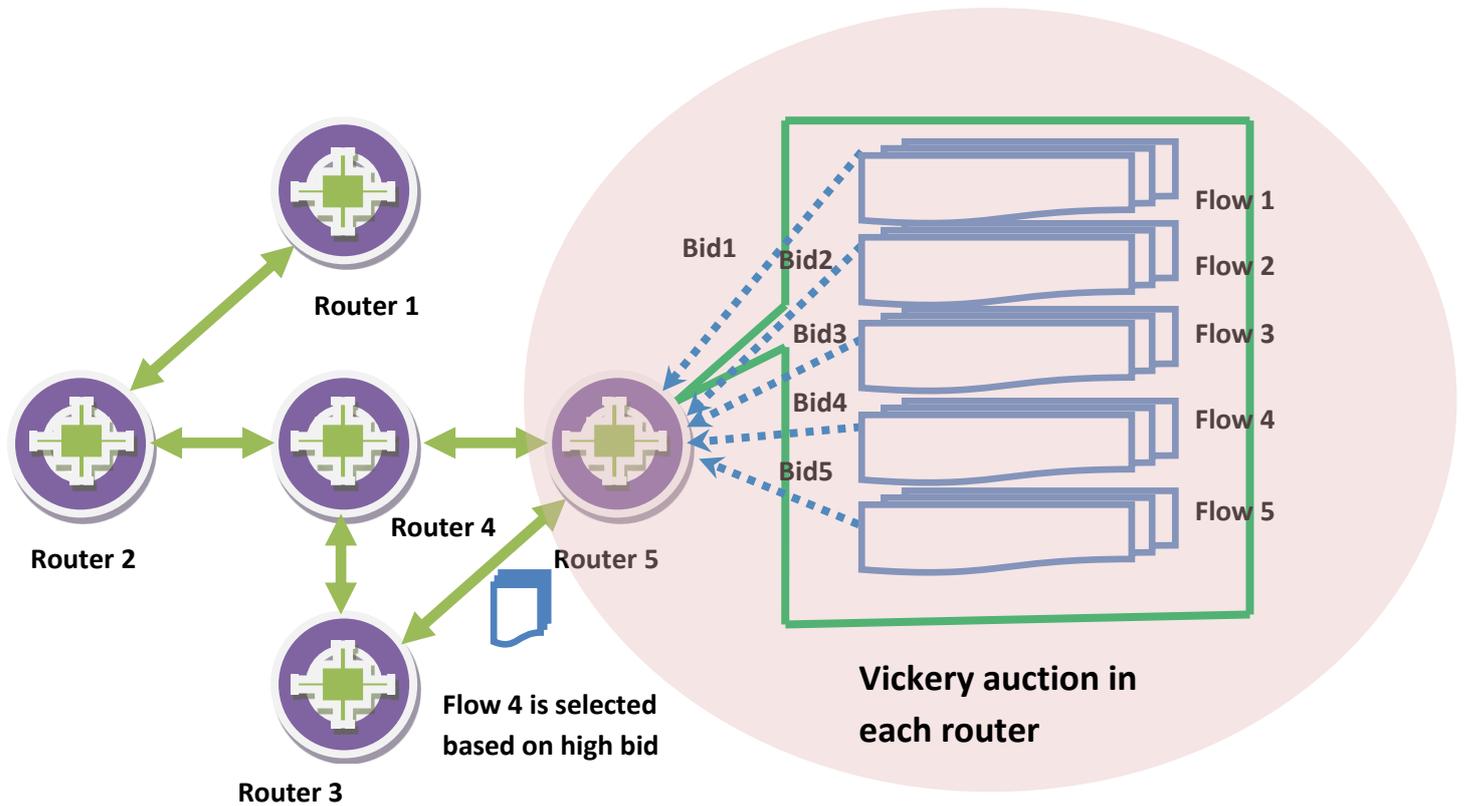

**Figure-5** *Auction based incentive scheme in packet forwarding*

***Router's Reserve Price:*** Reserve price for a router constitutes its operating cost, cost of facility, equipment and staffing etc. Therefore based on market research of utilities of different users routers can set their reservation cost to compensate for providing packet forwarding service.

***Maximum Bandwidth for a Flow:*** If requests of other flows are unchanged then a high bidder can get more bandwidth than others or it's the sum of all bandwidths allocated to low bidders or left-overs of high bidders.

Bottom line is in terms of mechanism, this scheme is quite simple as depicted in **Figure-5**; an auction process runs every time in the network to assign the price rate and bandwidth to the routers. Each packet flow constitutes a bid with it, that much amount of money he is agreed to pay. At each router a Vickery auction takes place for assigning different bandwidths to different packet flows based on their bids and it always converge in positive utility for the users.

# E-ARL: An Economic incentive scheme for Adaptive Revenue-Load-based dynamic replication of data in Mobile-P2P networks

Data availability and ease of access is elementary requirement and basic necessity for any network sharing system but it becomes more and more crucial and highly important when we have to deal with mobile networks such as mobile P2P networks since there network partitioning kind of situations may easily arrive due migration of some peer or some peer may get switched off or crashed due to less resources (battery, processing, storage etc.). Therefore proper replication of data is needed to avoid low data availability.

Incentive schemes for replication of data ensure adequate participation and collaboration of peers in the system. E-ARL suggests an incentive based economic data replication scheme which, not only helps to achieve better quality of service but also keeps track of load balancing and revenue balancing in the system and improves query response time significantly. On the hand this also takes care of free riders since they can't access anything without paying for their content.

***Basic E-ARL economic scheme***: Basic E-ARL economic scheme is quite intuitive in terms of processing, each data item is having some virtual monetary value based on its demand or importance let us say p when any mobile peer wants to access some data item that is kept by some other mobile peer then he has to pay p amount of virtual currency to the host peer. Requesting peer has to pay a commission (R) to those relay peers also who fall in its query path and perform relaying only. This means requesting peer has to pay total amount equal to

$$p + \sum_{i}^{r} R_i$$

Where r is number of relays happened to occur in successful query path. If requesting peer denies to pay this much amount than it's refrained from accessing the requested data time. Meeting query deadline is very important in such environment; as if requesting peer gets data item after query deadline then he is allowed to refuse to pay for his content even to relaying peers.

Revenue of a peer can be calculated as difference between virtual currencies earned by him against replica hosting or relaying and the amount spend by him for accessing desired content from other peers.

***Setting price of data items:*** One Intuitive policy could be that the more recently and frequently data item is requested more is the price but it's not considers the no of replicas of that particular item or number of peers serving that data item that should also be considered therefore as result price of any data item increases as it is served against requests originated from more mobile peers. Other than this replica consistency and faster query response can also be included when deciding price like more consistency level implies higher prices similarly more the faster response to the queries higher will be the prices for those data items.

The data items kept by hosts having low energy are priced high so that energy can be conserved and the low energy peers can be sustained in system. Similarly data items with high Time to live are priced higher enough because they can produce more revenue for longer time period. As replication of some data item increases its price decreases.

***E-ARL Bid Based Economic Replica Allocation:*** Replica allocation is done by a special peer called Super Peer (SP).

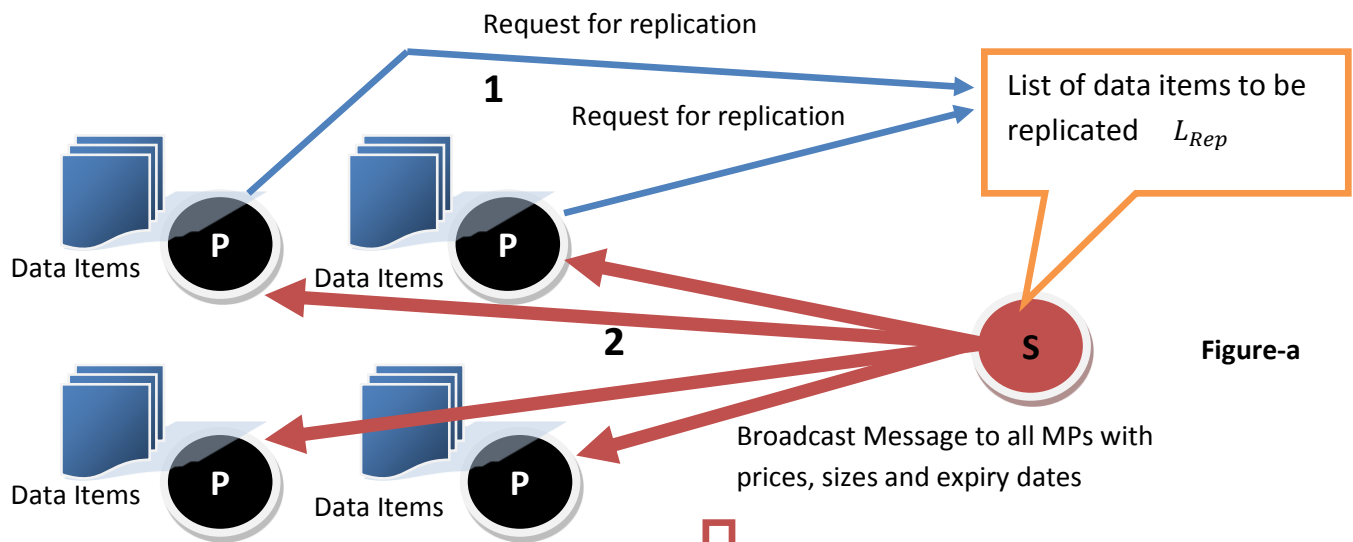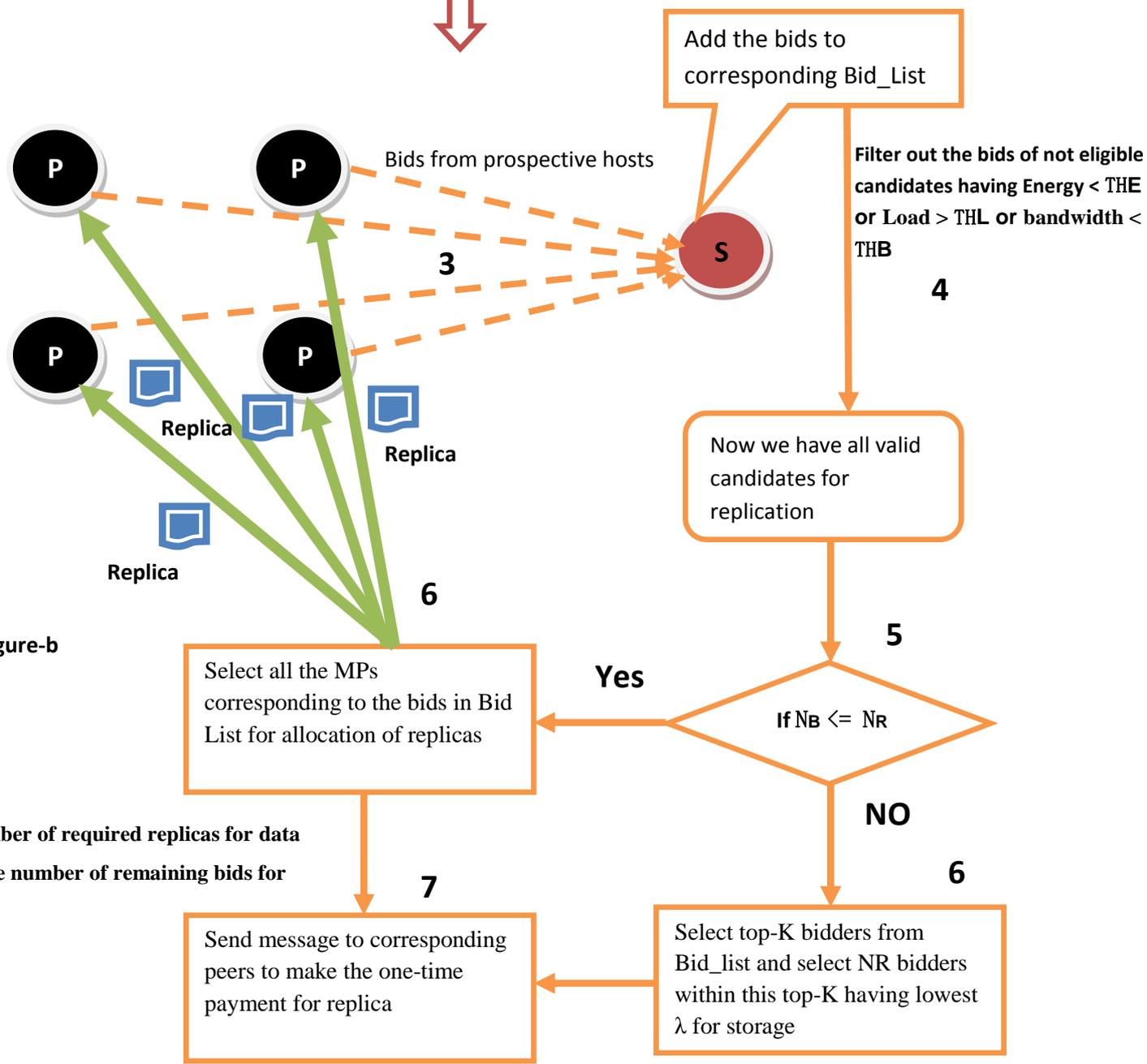

Figure-6 Replica Allocation in E-ARL

The peer who wants to provide his data for replication is called Provide peer and the peers who are interested in replicating data are called Host peers.

All the Host peers send their bids to SP and SP allocates replicas of higher priced data to higher bidders. Replicating data are called Host peers. All the Host peers send their bids to SP and SP allocates replicas of higher priced data to higher bidders.

***Basic Algorithm:*** we can divide whole algorithm into seven steps as depicted in **Figure-6** which is divided in two sub figures '**a**' and '**b**'.

***Step-1*** All the peers who want to provide their data items for bidding send their request to Super peer (SP). SP adds all the replication to a list called replica list ($L_{Rep}$).

***Step-2*** Now SP sends a broadcast message to all the peers for bidding and this broadcast messages entails the price, size and time to expire information for each data item in $L_{Rep}$. SP sends only one broadcast message for all items to conserve energy and bandwidth usage.

***Step-3*** After getting broadcast message from SP, all the interested peers send their bids for corresponding data items. SP adds all bids to a list called Bid List.

***Step-4*** SP traverses whole Bid List and filters out the not eligible bidders from the list. It eliminates all the bid entries whose energy is below the threshold **TH$_E$ or** whose load is above the threshold **TH$_L$ or** whose bandwidth is below the threshold **TH$_B$**.

***Step-5*** After getting list of eligible bids, SP calculates whether $N_B <= N_R$ is true or not. $N_R$ is the number of required replicas for data item d, $N_B$ is the number of remaining bids for data item d**.**

***Step-6*** If the inequality stated above comes to be true then SP simply allocates $N_R$ replicas to $N_R$ peers.

Else it selects top-K entries out of Bid_List and allocates $N_R$ replicas to peers having lowest λ for storage.

***Step-7*** Now SP sends messages to all peers who are allocated with replicas to release one-time payment for their replica.

Here k is usually set to **2\*$N_R$**.

We can see that this economic based replication scheme is fairly elegant but simple and bid based allocation of replicas is quite helpful in maintaining sufficient data availability and it also motivates peers to participate and earn wealth so that they can use it against their own data items needed. This scheme is also effective to reduce free riding.

# *Eco Rare: An Economic incentive scheme for Efficient Rare Data Accessibility in Mobile-P2P networks*

There can be such real world scenarios, where some rare data items get very important and demanded due to occurrence of some events otherwise they are rarely accessed and kept by very few peers. Eco Rare is specifically designed and devised for such scenarios.

We can think of some scenarios similar to as stated above, suppose there some gas emission happens in some city then there will be urgent need of gas masks in that particular location. Many users will now be interested in knowing about gas masks, their price and shops to purchase. Similar situation may arise in case of sudden snowfall, Then a huge number of people will be willing to buy shovels, will query about their prices, availability and probable locations to buy them. Information regarding these rare data is collected through advertisements published by shops (gas mask, shovel shops) but these kind of rare data is not accessible easily because of less replication of these data and which is quite likely because these data items are demanded only when certain event occurs.

***Basic Eco Rare:*** Each data item in this scheme is assumed to be associated with two kinds of prices namely

***Use_Price (Pu):*** This price serves only limited information about the data without any sort of ownership, therefore requester cannot resell this data to others. For an example let's take gas mask and shovels, in case of use price user gets only very little information for gas masks and shovels like some shops and prices of these goods at those shops.

***Sell_Price (Ps):*** Against the sell price user gets more detailed information about the data along with its ownership too, therefore user can now resell the data to others. A data item can have multiple owners in this scheme and due to this we can ensure availability of data even in the absence of some owner peers. In case of sell_price, user will get elaborated information for shopping gas masks and shovels like complete catalogues of more shops selling these items, how to purchase these items (online, phone delivery etc.) and even price comparison chart for different shops.

***Combatting with free riding:*** In this system requester have to pay relay charge to all relay peers, who are in successful query path so that quicker query forwarding can be done and in this way each mobile peer has to pay a constant money to relay peers to get his query forwarded. Due to these imposed costs free riding is almost impossible in such system because peers have to have currency to issue their own requests and to do so, they have to somehow earn money and they can only earn through either by hosting some data or relaying other peer's requests.

Eco Rare contributes in several ways to conventional peer to peer sharing system like

   **1)** It combats free riding effectively and in turn improves data availability as free riders are now actively involved.

   **2)** Since there can be multiple owners of a data item, multiple replicas of rare item are created.

   **3)** Availability and response time increases as relay peers also get incentives for relaying queries.

***Revenue of an MP:*** Revenue of a MP can be calculated as

Virtual currency earned by (Hosting data items +relaying service for others) **-** Virtual currency paid for (Buying data items + Relaying cost for own queries)

***Data selling mechanism of EcoRare:*** there are two kinds of selling ways defined by this scheme

***Query-based selling:*** In this case when a peer request to buy something, it's allocated with a replica of the corresponding data item means it becomes an owner of that item and it can resell it any time (**Figure-7 (a)**).

## Push/Pull-based selling:

**Push based:** In this case, the scenario where a peer is remained with very low energy. It wants to quit the system and sell his data items to get currency (**Figure-7 (b)**).

**Pull-based:** Pull based mechanism is usually adopted by free riders, since to issue query they must have some currency to pay and therefore they have buy some data from other peers so that they can earn money by selling them to others (**Figure-7 (c)**).

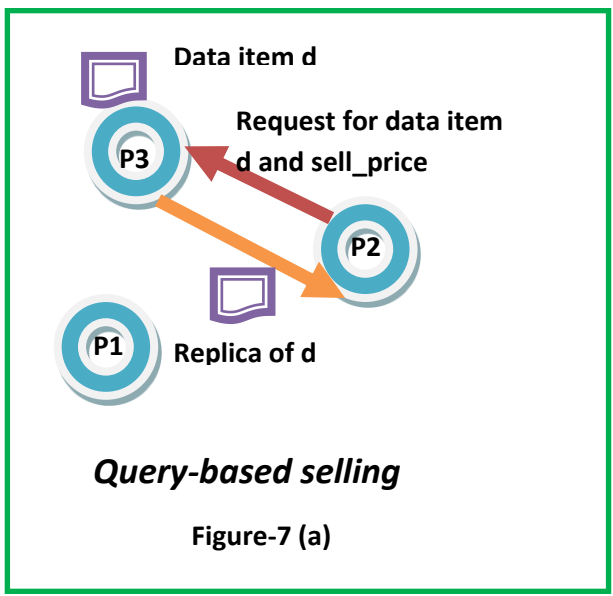

*Query-based selling*

Figure-7 (a)

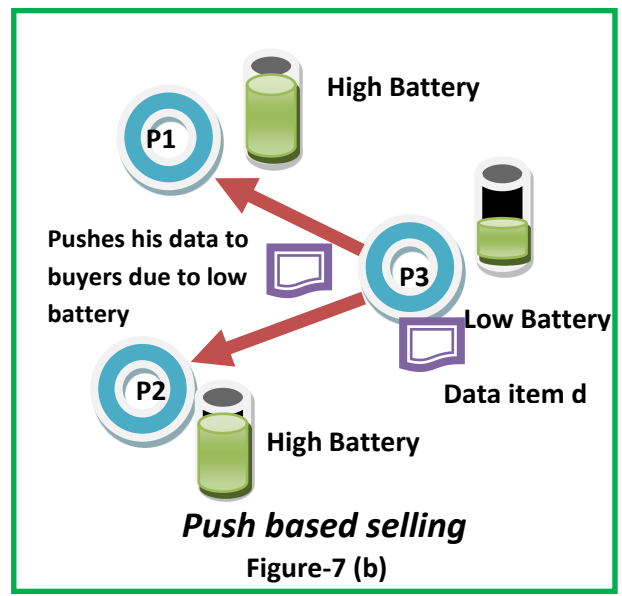

*Push based selling*

Figure-7 (b)

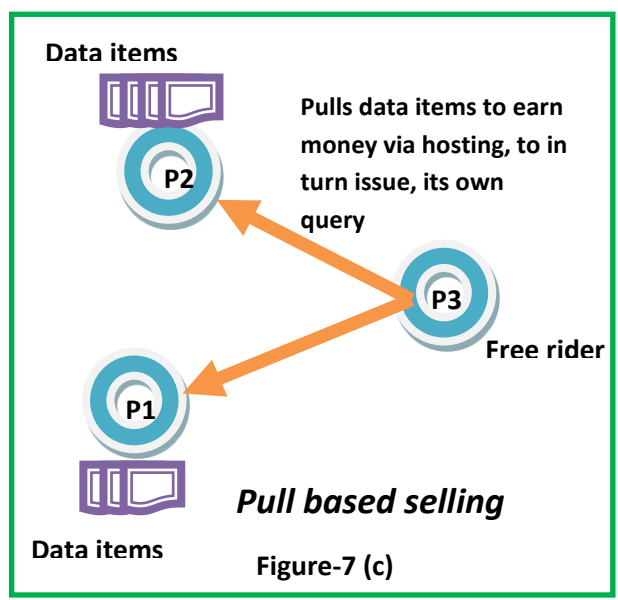

*Pull based selling*

Figure-7 (c)

***Algorithm EcoRare Seller MP:*** Data providing peers are called Sellers and query issuing peers are termed as Buyers. Steps for Eco Rare Seller algorithm are pretty straight forward.

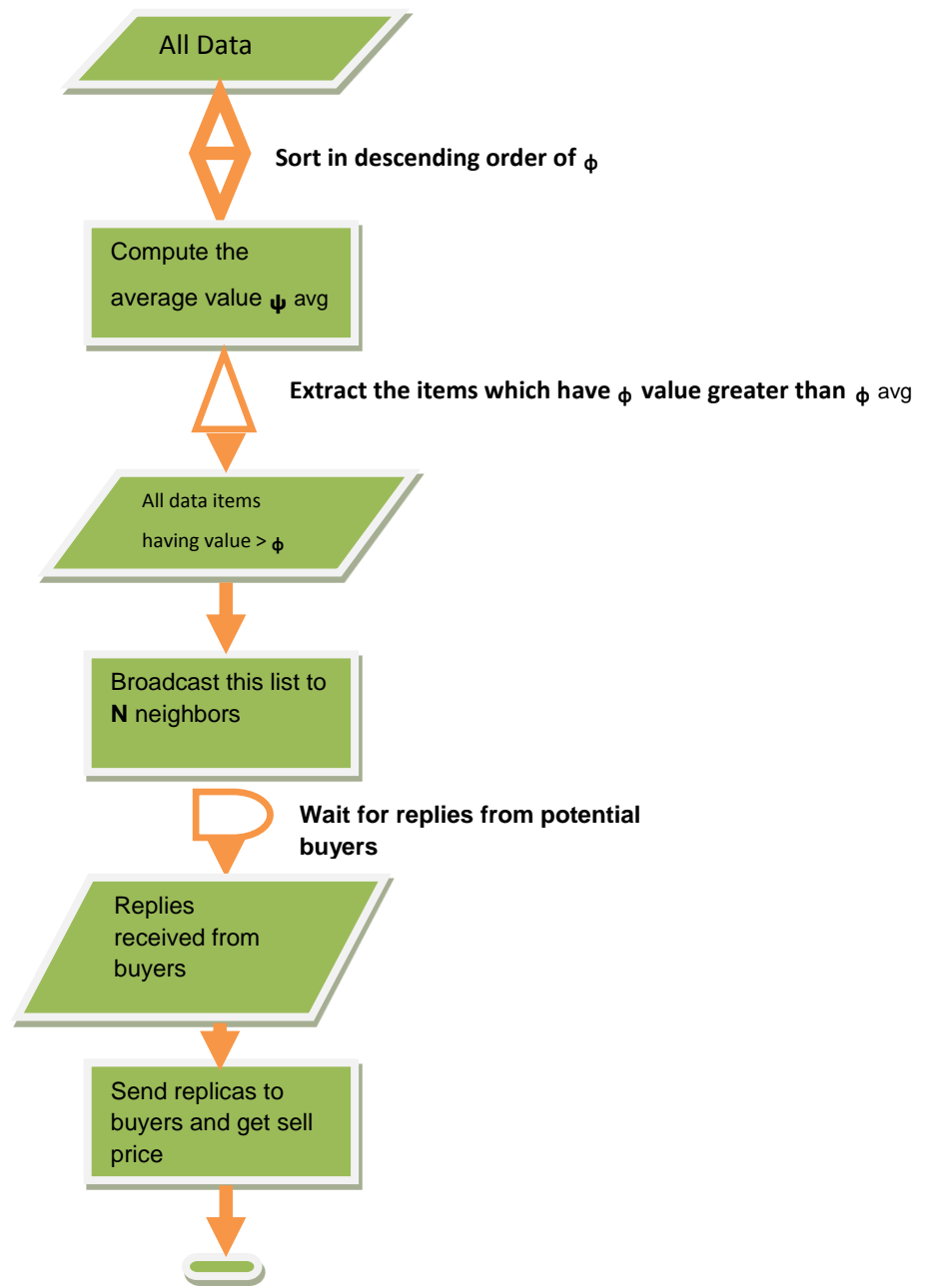

**Figure-8 Flow Chart for *Eco Rare Seller***

Note that $\phi$ signifies here the revenue-earning potential of data items and we have broadcasted the data items to only N neighbors rather than all of them to reduce communication overhead.

***Algorithm Eco Rare Buyer MP:*** Buyer algorithm is fairly intuitive and simple and it's depicted below via flow chart.

Storage space of buyer is denoted by $str_B$

Size of data is denoted by $Size_d$

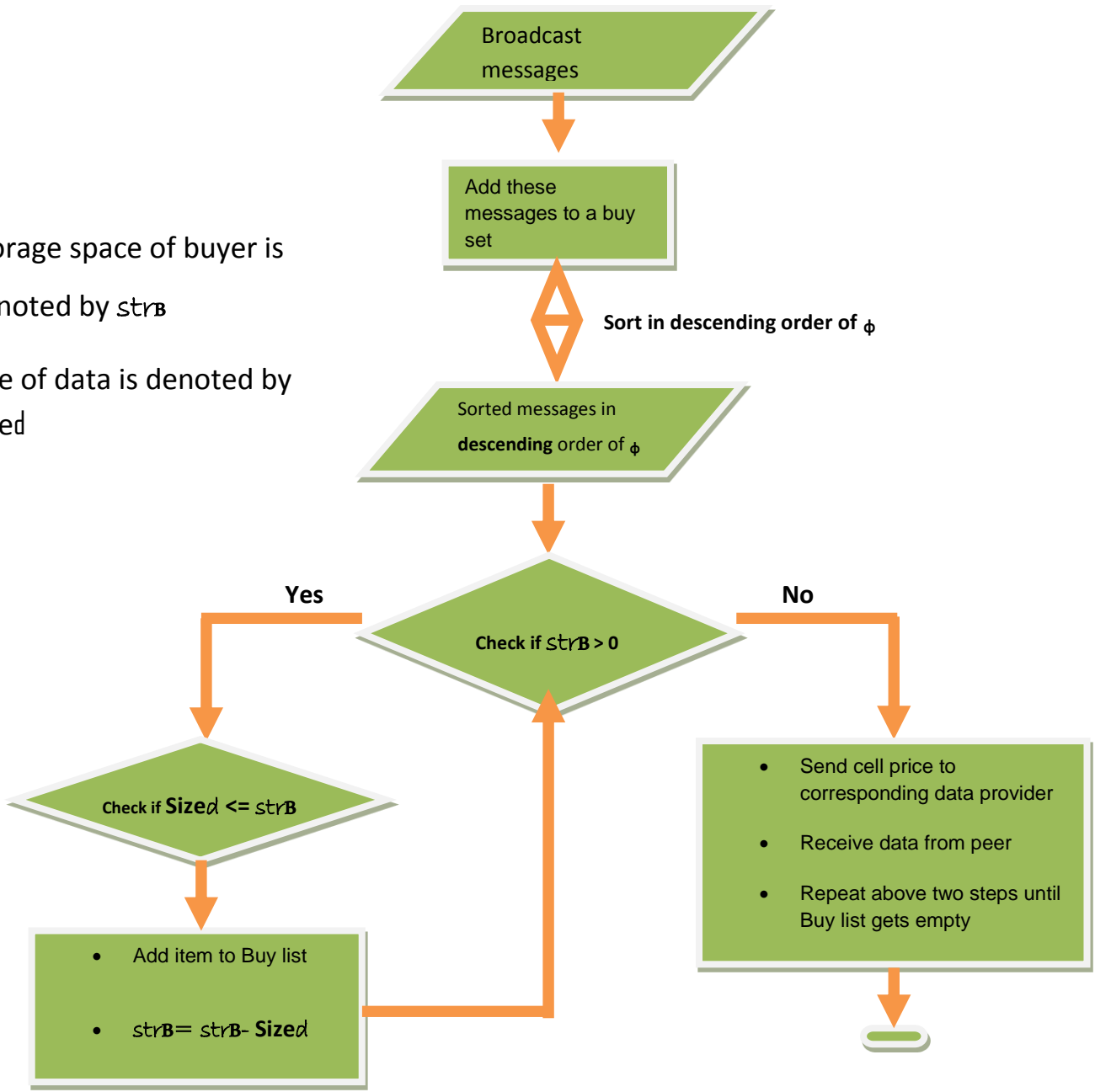

**Figure-9 Flow Chart for Eco Rare Buyer**

As a conclusion we can say that Eco Rare effectively combats free riding and involves free riders for participation and improves availability of rare data very efficiently.

## ABIDE: A Bid-based Economic Incentive Model for Enticing Non-cooperative Peers in Mobile-P2P Networks

In mobile peer to peer systems problem of "Tragedy of Commons" is very prevalent as we have discussed several times from the start of this survey. Incentive schemes are suggested to reduce the free riders and these schemes are proved to be fairly effective, ABIDE in essence is a similar effort to overcome the free riding problem by a novel bid based incentive model.

It not only motivates relay peers to work as broker for value added routing but also helps new peers to earn revenues in order to obtain services and encourages effective data sharing and resource sharing.

***Basic model of ABIDE:*** Each Mobile peers creates an index of services (data item + peers having this data) and each relay peer is provided with an amount of incentive for searching on behalf of query issuer's query. These indexes are generated on the fly based queries issued to it and different peers may have entirely different indexes.

User issues his query through broadcasting, the peer who receives this query checks whether needed data is indexed in its index against some peer or not and if it happens to be false, it just forwards the query to some neighbor peer to get relay commission. On the other hand if the result is indexed then it issues a new query to destined peer to get answer and it acts as a broker. Commission for relaying is quite less than brokering.

*Network topology in ABIDE:*

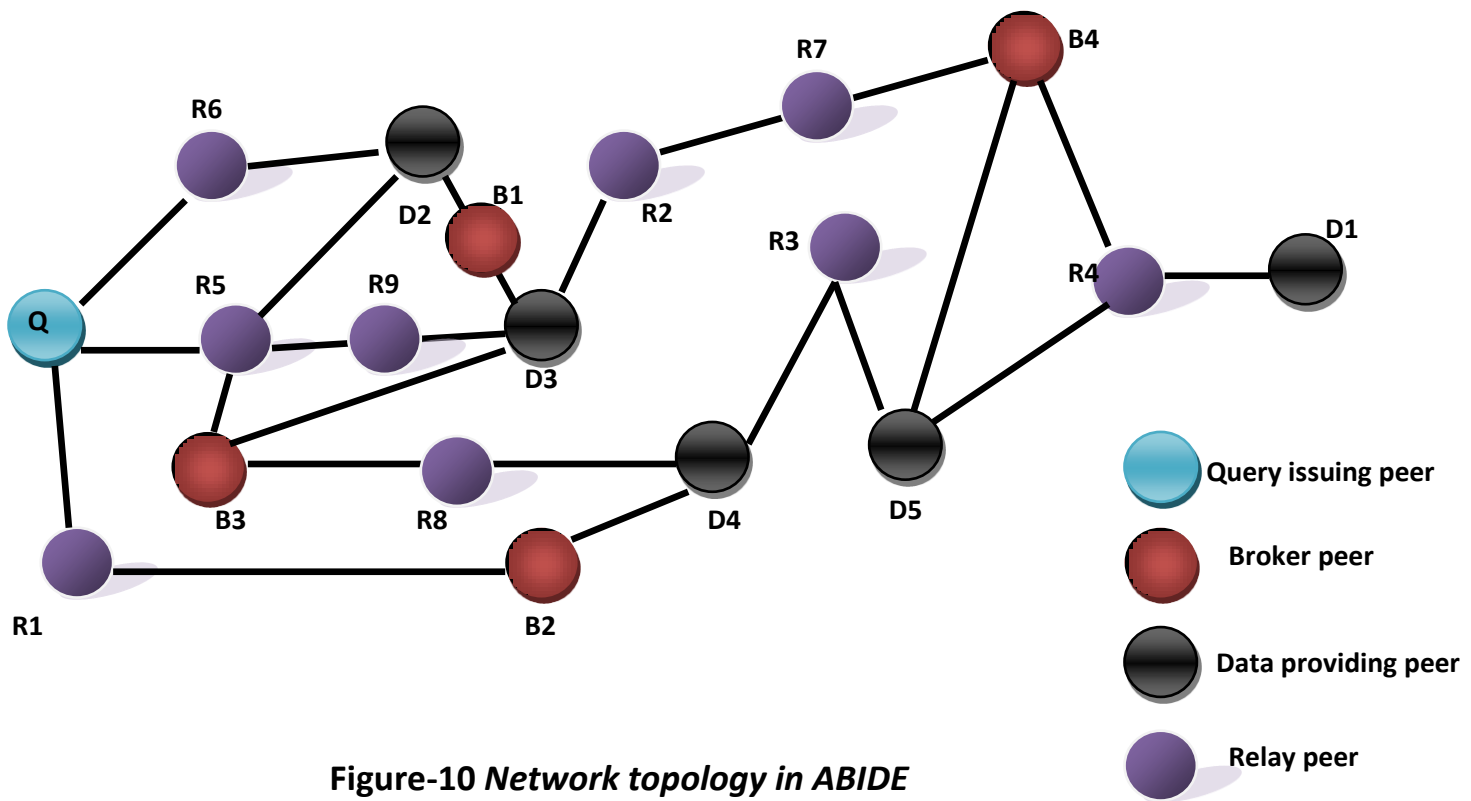

**Figure-10** *Network topology in ABIDE*

In **figure-10** Q is query issuing peer and from R1 to R9 are relay peers similarly from B1 to B4 are the broker peer and from D1 to D5 are the data providing peers. Let's take some scenarios and analyse what will be the role of different entries as depicted above.

**Scenario1:** Suppose Q issues a query and the desired data is at D2 data providing peer. In this case we can see that only one relay peer is there between Q and D2. Therefore Q has to pay 1 unit relay cost to R6 and price of data to D2.

**Scenario2:** Suppose Q issues a query for let's say data d and data d have two copies in system i.e. at D4, D3. Now there are three promising query paths to get data d. First is {R5, R9, D3}, second is {R5, B3, R8, D4} and third is {R1, B2, D4}. First and second path is having two relay peers but second is having one broker peer too that's why we can omit second path for consideration. Now only first and third are left for candidature. First path is having two relays and third path is having one relay and one broker, since relay charge is less than broker's charge therefore path first will be chosen (assuming issuer has index for required data himself).

***Algorithm ABIDE Query Issuing MPs:*** Steps in algorithm

**Steps1:** Broadcast query for desired data item d.

**Steps2:** Receive all the bids and evaluate value of $\gamma$ each bid.

**Steps3:** Select the bid for which $\gamma$ is highest and select the corresponding broker and send message to him.

**Steps4:** receive data from broker and his commission based strategy chosen (PPA or NPA).

Where PPA stands for *Privacy-Preserving Auction model where* query issuer remains anonymous to data provider and NPA stands for *Non-Privacy preserving Auction model* where anonymity is not maintained. Where $\gamma$ is a score which constitutes of two parameters namely query response and data quality.

$$\gamma = a \times RT + b \times DQ$$

**RT** is response time for query and **DQ** is Data quality. a and b are constants and calculated experimentally.

## *Conclusion*

Firstly, we have covered several incentive based schemes suggested for mobile peer to peer networks having different approaches to tackle similar problems and even we have gone through entirely different scenarios and problems of modern P2P systems. Mobile peer to peer environment, which is quite constrained in terms of energy, resources and protocols, we have seen these techniques are provably beneficial for such unstable mobile environment which can get easily posed to unpredictable situations like network partitioning, peer quitting and critically less data availability.

Secondly, we have described what the free riding problem is all about and how to handle free riding effectively in mobile peer to peer environment. Since free riding is very challenging for P2P systems, we have elaborated this issue quite well and described handful schemes to handle it. We have also described some schemes for data replication for MP2P environment so that data availability can be improved and even to increase availability of some particular kind of data only, whose demand arises based on occurrence of some specific events.

As a conclusion, incentive schemes for Mobile Peer–to-Peer Systems are fairly effective and promising to solve challenges of modern P2P system and free riding problem.